\documentstyle[prl,preprint,aps]{revtex}
\begin{document}
\draft
\title{Gravitational-Radiation Damping of
Compact Binary Systems\\ to Second Post-Newtonian order}
\author{Luc Blanchet$^1$, Thibault Damour$^{2,1}$, Bala R. Iyer$^3$,
Clifford M. Will$^4$,\\
and Alan G. Wiseman$^5$\cite{alan}}
\address{$^1$D\'epartement d'Astrophysique Relativiste et de
Cosmologie\cite{cnrs},
Observatoire de Paris, 92195 Meudon Cedex, France}
\address{$^2$Institut des Hautes \'Etudes Scientifiques,
91440 Bures sur Yvette, France}
\address{$^3$Raman Research Institute, Bangalore 560 080, India}
\address{$^4$McDonnell Center for the Space Sciences,
Department of Physics, Washington University,\\
St. Louis, Missouri 63130}
\address{$^5$Department of Physics and Astronomy,
Northwestern University, Evanston, Illinois 60208}
\date{\today}
\maketitle
\widetext
\begin{abstract}
The rate of gravitational-wave energy loss from inspiralling binary
systems of compact objects of arbitrary mass is derived through second
post-Newtonian (2PN) order $O[(Gm/rc^2)^2]$ beyond the quadrupole
approximation.  The result has been derived by two independent
calculations of the (source) multipole moments. The 2PN terms,
and in particular the finite mass contribution therein (which
cannot be obtained in perturbation calculations of black hole
spacetimes), are shown to make a significant contribution to the
accumulated phase of theoretical templates to be used in matched
filtering of the data from future gravitational-wave detectors.
\end{abstract}

One of the most promising astrophysical sources of gravitational
radiation for detection by large-scale laser-interferometer systems
as the US LIGO or the French-Italian VIRGO projects \cite{ligo} is the
inspiralling compact binary.  This is a binary system of neutron stars
or black holes whose orbit is decaying toward a final coalescence
under the dissipative effect of gravitational radiation reaction.
For much of the evolution of such systems, the gravitational waveform
signal is an accurately calculable ``chirp'' signal that sweeps in
frequency through the detectors' sensitive bandwidth, typically between
10~Hz and 1000~Hz.  Estimates of the rate of such inspiral events range
from 3 to 100 per year, for signals detectable out to hundreds of Mpc
by the advanced version of LIGO\cite{clark}.

In addition to outright detection of the waves, it will be possible
to determine parameters of the inspiralling systems, such as the masses
and spins of the bodies\cite{jugger,finn,cutler}, to measure cosmological
distances\cite{cosmo}, to probe the non-linear regime of general
relativity\cite{sathya}, and to test alternative gravitational
theories\cite{brans}.  This is made possible by the technique of
matched filtering of theoretical waveform templates, which depend
on the source parameters, against the broad-band detector
output\cite{filters}.

Roughly speaking, any effect that causes the template to differ from
the actual signal by one cycle over the 500 to 16,000 accumulated cycles
in the sensitive bandwidth will result in a substantial reduction in
the signal-to-noise ratio.  This necessitates knowing the prediction of
general relativity for gravitational radiation damping, and its effect
on the orbital phase, to substantially higher accuracy than that provided
by the lowest-order quadrupole, or Newtonian approximation.  If
post-Newtonian corrections to the quadrupole formula scale as powers
of $v^2 \approx m/r$ ($G=c=1$), then, say, for a double neutron-star
inspiral in the LIGO/VIRGO bandwidth, with $m/r$ typically around
$10^{-2}$, corrections {\it at least} of order
$(m/r)^2 \sim 10^{-4}$ will be needed in order to be accurate
to one cycle out of the 16,000 cycles accumulated for this process.
This corresponds to corrections of {\it second} post-Newtonian (2PN)
order.

Although numerous corrections to the quadrupole energy-loss formula
have previously been calculated (for a summary, see\cite{nishi}),
the 2PN contributions to the energy loss rate for arbitrary masses
have not been derived.  This paper presents these contributions for
the first time, discusses their significance for gravitational-wave
data analysis, and outlines the derivation.

The central result is this:  through 2PN order, the rate of energy loss,
$dE/dt$, from a binary system of two compact bodies of mass $m_1$
and $m_2$, orbital separation $r$, and spins ${\bf S}_1$ and
${\bf S}_2$, in a nearly circular orbit (apart from the adiabatic
inspiral) is given by
\begin{eqnarray}
    -{dE \over dt} =
&& {32 \over 5} \eta^2 {\left (m \over r \right )}^5
   \biggl [ 1-{m \over r} \left ( {2927 \over 336}
   + {5 \over 4} \eta \right ) \nonumber \\
&& +\, {\left (m \over r \right )}^{3/2}
     \biggl \{ 4\pi - {1 \over 12} \Sigma_i
     \left ( 73 {m_i^2 \over m^2} + 75 \eta \right )
     {\bf \hat L} \cdot \mbox{\boldmath$\chi$}_i \biggr \} \nonumber \\
&& +\, {\left (m \over r \right )}^2 \biggl
     \{ {293383 \over 9072} + {380 \over 9} \eta \nonumber \\
&& -\, {\eta \over 48} \left ( 223 \mbox{\boldmath$\chi$}_1
     \cdot \mbox{\boldmath$\chi$}_2 - 649 {\bf \hat L}
     \cdot \mbox{\boldmath$\chi$}_1 {\bf \hat  L}
     \cdot \mbox{\boldmath$\chi$}_2 \right ) \biggr \} \biggr ] \,,
\label{edot}
\end{eqnarray}
where $\bf \hat L$ is a unit vector directed along the orbital angular
momentum, $m = m_1 + m_2$, $\eta=m_1m_2/m^2$,
$\mbox{\boldmath$\chi$}_1={\bf S}_1/m_1^2$,
$\mbox{\boldmath$\chi$}_2={\bf S}_2/m_2^2$ and $\Sigma_i$ denotes the
sum over ${i=1,2}$.  The terms in square brackets in Eq.~(\ref{edot})
are respectively:  at lowest order, Newtonian (quadrupole);  at order
$m/r$, 1PN \cite{wagwill};  at order $(m/r)^{3/2}$, the non-linear
effect of ``tails'' ($4\pi$ term) \cite{poissona,tails},
and spin-orbit effects\cite{kww,kidder};  and at order $(m/r)^2$,
the 2PN terms (new with this paper), and spin-spin
effects\cite{kww,kidder,spincomment}.

For the special case of a test mass orbiting a massive black hole,
perturbation theory has been used to derive an analogous analytic
formula (apart from spin-spin effects)\cite{poissona,poissonb},
and for non-rotating holes, to extend the expansion through
the equivalent of 4PN order\cite{tagoshi}.  The test-body $\eta=0$
limit of Eq.~(\ref{edot}) agrees completely with these results to
the corresponding order.

The equations of motion for circular orbits, correct to 2PN order including
spin effects, yield for the orbital angular velocity $\omega \equiv v/r$ and
the orbital energy $E$~\cite{dds,kww,nishi}
\begin{mathletters} \label{motion}
\begin{eqnarray}
  \omega^2 =
&& {m \over r^3} \biggl[ 1-{m \over r}(3- \eta )
    -\left( {m \over r}\right)^{3/2}
    \Sigma_i \left( 2{m_i^2 \over m^2} + 3 \eta \right )\nonumber \\
&&\quad \times\, {\bf \hat L} \cdot \mbox{\boldmath$\chi$}_i
   +\left( {m \over r} \right)^2
   \biggl\{ 6 + {41 \over 4} \eta + \eta^2 \nonumber \\
&& \quad -\, {3 \eta \over 2}
   \left( \mbox{\boldmath$\chi$}_1\cdot
          \mbox{\boldmath$\chi$}_2 -3{\bf \hat L} \cdot
          \mbox{\boldmath$\chi$}_1 {\bf \hat L}
   \cdot \mbox{\boldmath$\chi$}_2 \right ) \biggr\} \biggr ] \,,\\
   -E =
&& \eta {m^2 \over 2r}
   \biggl [ 1-{1 \over 4}{m \over r} ( 7- \eta )
  +\left({m \over r} \right)^{3/2}
   \Sigma_i \left( 2{m_i^2 \over m^2} + \eta \right)\nonumber \\
&&\qquad\times\, {\bf \hat L} \cdot \mbox{\boldmath$\chi$}_i
  +\left({m \over r} \right)^2
   \biggl\{ {1 \over 8} ( -7 + 49 \eta + \eta^2 ) \nonumber \\
&& \qquad +\, {\eta \over 2} \left ( \mbox{\boldmath$\chi$}_1
   \cdot \mbox{\boldmath$\chi$}_2 - 3 {\bf \hat L}
   \cdot \mbox{\boldmath$\chi$}_1 {\bf \hat L}
   \cdot \mbox{\boldmath$\chi$}_2 \right) \biggr \} \biggr ] \ .
\end{eqnarray}
\end{mathletters}
Combining
Eqs.~(\ref{edot}) and (\ref{motion}), one can express the rate of
change $\dot\omega$ of the angular velocity as a function of $\omega$,
and get
\begin{eqnarray}  \label{omegadot}
     \dot\omega
&=& {96 \over 5} \eta m^{5/3} \omega^{11/3}
    \biggl [ 1-\left( {743 \over 336}
           +{11 \over 4} \eta \right) (m\omega)^{2/3} \nonumber \\
&& +\, (4 \pi - \beta) (m \omega ) \nonumber \\
&& + \left ( {34103 \over 18144} + {13661 \over 2016} \eta
     +{59 \over 18} \eta^2 + \sigma\right )
      (m\omega )^{4/3} \biggr ] \,,
\end{eqnarray}
where the spin-orbit ($\beta$) and spin-spin ($\sigma$) parameters are
given by
$\beta ={1\over 12}\Sigma_i (113{m_i^2/m^2}+75\eta ){\bf \hat L}
\cdot \mbox{\boldmath$\chi$}_i$ , and
$\sigma = (\eta /{48})(-247 \mbox{\boldmath$\chi$}_1
\cdot \mbox{\boldmath$\chi$}_2 + 721 {\bf \hat L}
\cdot \mbox{\boldmath$\chi$}_1 {\bf \hat L} \cdot
\mbox{\boldmath$\chi$}_2 )$.
 From that one calculates the accumulated number of gravitational-wave
cycles ${\cal N} =\int (f/\dot f)df$, where $f=\omega/\pi$ is the
frequency of the quadrupolar waves, in terms of the frequencies at
which the signal enters and leaves the detectors' bandwidth.
In order to avoid complications caused by spin-induced precessions
of the orbital plane\cite{aposto,kidder}, we assume that the spins
are aligned parallel to the orbital angular momentum (in particular
$\beta$ and $\sigma$ remain constant).  Using 10~Hz as the entering
frequency of LIGO/VIRGO-type detectors, set by seismic noise, and,
as the exit frequency, the smaller of either 1000~Hz (set by
photon-shot noise) or the frequency corresponding to the innermost
stable circular orbit and the onset of plunge [for small mass ratio,
$f_{\max} = 1/({6}^{3/2} \pi m)$\cite{plunge}], we find contributions
to the total number of observed wave cycles from the various
post-Newtonian terms listed in Table~1.

Because $\chi \le 1$ for black holes, and $\lesssim 0.63-0.74$ for
neutron stars (depending on the equation of state, see~\cite{NS}),
$\beta$ and $\sigma$ are always less than $\sim 9.4$ and $\sim 2.5$,
respectively.  However, if we consider models for the past and future
evolution of observed binary pulsar systems such as PSR 1534+12 and
PSR 1913+16, we find (using a conservative upper limit for moments
of inertia) that
$\chi_1^{1534+12} < 5.2\times 10^{-3}$,
$\chi_1^{1913+16} < 6.5 \times 10^{-3}$, and we expect
$\chi_2 \lesssim 7 \times 10^{-4}$.  If such values are typical,
both the spin-orbit and (a fortiori) the spin-spin terms will make
negligible contributions to the accumulated phase.

Table 1 demonstrates that the 2PN terms, and notably the finite-mass
($\eta$-dependent) contributions therein (which cannot be obtained
by test-body approaches), make a significant contribution to the
accumulated phase, and thus must be included in theoretical templates
to be used in matched filtering.  The additional question of how the
presence of these terms will affect the accuracy of estimation of
parameters in the templates can only be answered reliably using a full
matched filter analysis~\cite{finn,cutler}.  This is currently in
progress~\cite{poissonwill}.

The remainder of this paper outlines the derivations leading to this
result.  Two entirely independent calculations were carried out,
using different approaches, one by BDI, using their previously
developed generation formalism~\cite{bdirefs,luc}, the other by WW,
using a formal slow-motion expansion originated by Epstein and
Wagoner~\cite{ew}.  Details of these calculations will be published
elsewhere~\cite{2papers}.

Both approaches begin with Einstein's equations written in harmonic
coordinates (see \cite{luc} for definitions and notation).  We define
the field $h^{\alpha\beta}$, measuring the deviation of the ``gothic''
metric from the Minkowski metric
$\eta^{\alpha\beta}={\rm diag}(-1,1,1,1)$:
$h^{\alpha\beta}=\sqrt{-g} g^{\alpha\beta}-\eta^{\alpha\beta}$.
(Greek indices range from 0 to 3, while Latin range from 1 to 3).
Imposing the harmonic coordinate condition
$\partial_\beta h^{\alpha\beta}=0$ then leads to the field equations
\begin{equation}
   \Box h^{\alpha\beta}= 16\pi (-g) T^{\alpha\beta}
 + \Lambda^{\alpha\beta}(h)
   \equiv 16\pi\tau^{\alpha\beta} \, ,
\label{relaxed}
\end{equation}
where $\Box$ denotes the flat spacetime d'Alembertian operator,
$T^{\alpha\beta}$ is the matter stress-energy tensor, and
$\Lambda^{\alpha\beta}$ is an effective gravitational source containing
the non-linearities of Einstein's equations.  It is a series in powers
of $h^{\alpha\beta}$ and its derivatives; both quadratic and cubic
nonlinearities in $\Lambda^{\alpha\beta}$ play an essential role
in our calculations.

{\it Post-Minkowski matching approach (BDI)}.  This approach proceeds
through several steps.  The first consists of constructing an iterative
solution of Eq.~(\ref{relaxed}) in an inner domain (or near zone)
that includes the material source but whose radius is much less than
a gravitational wavelength.  Defining source densities
$\sigma =T^{00}+T^{kk}$, $\sigma_i=T^{0i}$, $\sigma_{ij}=T^{ij}$,
and the retarded potentials $V=-4\pi\Box_R^{-1}\sigma$,
$V_i =-4\pi\Box_R^{-1}\sigma_i$, and
$W_{ij}=-4\pi\Box_R^{-1}[\sigma_{ij}+(4\pi)^{-1}
(\partial_i V\partial_j V-{1\over 2} \delta_{ij}
\partial_k V \partial_k V)]$,
where $\Box_R^{-1}$ denotes the usual flat spacetime retarded
integral, one obtains the inner metric $h_{\rm in}^{\alpha\beta}$
to some intermediate accuracy $O(6,5,6)$:
$h^{00}_{\rm in}=-4V+4(W_{ii}-2V^2)+O(6)$,
$h^{0i}_{\rm in}=-4V_i+O(5)$,
$h^{ij}_{\rm in}=-4W_{ij}+O(6)$, where $O(n)$ means a term of order
$\varepsilon^n$ in the post-Newtonian parameter $\varepsilon\sim v/c$.
{}From this, one constructs the inner metric with the higher accuracy
$O(8,7,8)$ needed for subsequent matching as
$h_{\rm in}^{\alpha\beta}=\Box_R^{-1}
[16\pi\overline{\tau}^{\alpha\beta}(V,W)]+O(8,7,8)$,
where $\overline{\tau}^{\alpha \beta} (V,W)$
denotes the right-hand-side of Eq.~(\ref{relaxed}) when retaining
all the quadratic and cubic nonlinearities to the required
post-Newtonian order in the near zone, and given as explicit
combinations of derivatives of $V,V_i$ and $W_{ij}$.  The second
step consists of constructing a generic solution of the {\it vacuum}
Einstein equations (Eq.~(\ref{relaxed}) with $T^{\alpha\beta} = 0$),
in the form of a multipolar-post-Minkowskian expansion that is valid
in an external domain which overlaps with the near zone and extends
into the far wave zone.  The construction of $h^{\alpha \beta}$ in
the external domain is done algorithmically as a functional of a set
of parameters, called the ``canonical'' multipole moments
$M_{i_1 \ldots i_l} (t)$, $S_{i_1 \ldots i_l} (t)$ which are symmetric
and trace-free (STF) Cartesian tensors.  Schematically,
$h_{\rm ext}^{\alpha \beta} = {\cal F}^{\alpha\beta} [M_L ,S_L]$
where $L\equiv i_1 \ldots i_l$ and where the functional dependence
includes a non-local time dependence on the past ``history"
of $M_L (t)$ and $S_L (t)$.  The third, ``matching'' step consists
of requiring that the inner and external metrics be equivalent
(modulo a coordinate transformation) in the overlap between the
inner and the external domains.  This requirement determines the
relation between the canonical moments and the inner metric
(itself expressed in terms of the source variables).  Performing
the matching through 2PN order~\cite{luc} thus determines
$M_L (t)=I_L [\overline{\tau}^{\alpha \beta}]+O(5)$,
$S_L (t) = J_L [\overline{\tau}^{\alpha \beta}] + O(4)$,
where the ``source'' moments $I_L$ and $J_L$ are given by some
mathematically well-defined (analytically continued) integrals
of the quantity $\overline{\tau}^{\alpha\beta} (V,W)$ which appeared
as source of $h_{\rm in}^{\alpha \beta}$.  When computing the source
moments we neglect all finite size effects, such as spin (which to 2PN
accuracy can be added separately) and internal quadrupole effects.
The final result for the 2PN quadrupole moment reads, in the case
of a {\it circular} binary,
\begin{equation} \label{quadrupole}
  I_{ij} = {\rm STF}_{ij}\, \eta m[Ar^{i}r^{j}+Br^2v^{i}v^{j}] \,,
\end{equation}
with $A=1-{1\over 42}(m/r)(1+39\eta )-{1\over 1512}(m/r)^2
(461+18395\eta +241\eta^2)$,
$B={11\over 21}(1-3\eta )+{1\over 378}(m/r)(1607-1681\eta +229\eta^2)$.
The final step consists of computing from the external metric
$h_{\rm ext}^{\alpha\beta}$ the gravitational radiation emitted
at infinity. This entails introducing a (non harmonic) ``radiative''
coordinate system $X^{\mu} = (T,X^i)$ adapted to the fall-off of the
metric at future null infinity. The transverse-traceless (TT)
asymptotic waveform $h_{ij}^{TT}$ can be uniquely decomposed
into two sets of STF ``radiative'' multipole moments $U_L$, $V_L$
which are then computed as some non linear functionals of the
canonical moments, and therefore of the source multipole moments.
For instance, up to $O(5)$,
\begin{equation} \label{tail}
  U_{ij}(T) = I_{ij}^{(2)} (T) + 2m\int_{-\infty}^{T} dT' \ln
  \left( {T-T' \over {2b_1}}\right) I_{ij}^{(4)} (T') \,,
\end{equation}
which contains a non-local ``tail" integral (in which
$b_1=be^{-11/12}$ where $b$ is a freely specifiable parameter
entering the coordinate transformation
$x^{\mu}\rightarrow X^{\mu}:T-R=t-|{\bf x} |-2m\ln (|{\bf x}|/b)$).
The superscript $(n)$ denotes $n$ time derivatives.
The energy loss is given by integrating the square of
$\partial h_{ij}^{TT} / \partial T$ over the sphere at infinity.
At 2PN order this leads to
\begin{eqnarray} \label {edotbdi}
   -{dE \over dT}
&=& {1\over 5} U_{ij}^{(1)} U_{ij}^{(1)} + {1\over 189}
    U_{ijk}^{(1)} U_{ijk}^{(1)}
    +{16 \over 45} V_{ij}^{(1)} V_{ij}^{(1)}\nonumber\\
&&+ {1\over 9072} U_{ijkm}^{(1)} U_{ijkm}^{(1)} + {1\over 84}
    V_{ijk}^{(1)} V_{ijk}^{(1)} \ .
\end{eqnarray}
Inserting the 2PN expression~(\ref{quadrupole}) of the source
quadrupole into the radiative quadrupole~(\ref{tail}),
and using the previously derived 1PN expressions for the
other multipole moments, we end up with the energy loss~(\ref{edot}).

{\it Epstein-Wagoner approach (WW)}.  This approach starts by
considering\\
$h^{\alpha\beta}=\Box_R^{-1}(16\pi\tau^{\alpha\beta})$
as a formal solution of Eq.~(\ref{relaxed}) everywhere.
One then expands the retarded integral $\Box_R^{-1}$ to leading
order in $1/R$ in the far zone, while the retardation is expanded
in a slow-motion approximation.  Using identities, such as
$\tau^{ij}={1 \over 2}\partial^2(\tau^{00}x^ix^j )/\partial t^2+$
spatial divergences, which result from
$\partial_\beta \tau^{\alpha\beta}=0$
(a consequence of the harmonic gauge condition), we express the
radiative field as a sequence of Epstein-Wagoner multipole moments,
\begin{equation}
  h^{ij}_{TT} = {2 \over R}{d^2 \over {dt^2}}
  \sum_{m=0}^\infty n_{k_1}
  \dots n_{k_m} I_{\rm EW}^{ijk_1 \dots k_m} (t-R) _{TT} \ ,
\label{hew}
\end{equation}
where $I_{\rm EW}^{ijk_1 \dots k_m}$ are integrals over space of moments
of the source $\tau^{\alpha\beta}$ (e.g. $I_{\rm EW}^{ij}=\int \tau^{00}
x^ix^j d^3x$; see \cite{ew,magnum} for formulae).
To sufficient accuracy for the radiative field, the 2-index moment
must be calculated to 2PN order, the 3 and 4-index moments to PN order,
and the 5 and 6-index moments to Newtonian order.  The moments of the
compact-support source distribution $(-g)T^{\alpha \beta}$ are
straightforward.  Contrary to what happens in the BDI calculation
where the matching leads to mathematically well-defined formulas
for the source multipole moments, the EW moments of the non-compact
$\Lambda^{\alpha \beta}$ source are given by formally divergent
integrals.  To deal with this difficulty we define a sphere of
radius ${\cal R} \gg \lambda \sim r/\varepsilon$ centered on the
center of mass of the system, and integrate the non-compact moments
within the sphere.  Many integrations by parts are carried out
to simplify the calculations, and the resulting surface terms
are evaluated at $\cal R$ and kept.  The divergent terms are
proportional to $\cal R$, and signal the failure of the slow-motion
expansion procedure extended into the far zone.  We discard the
divergent terms.  In order to compare directly with BDI, we
transform the EW moments into STF moments using the projection
integrals given by Thorne~\cite{rmp}.  For the quadrupole moment,
for instance, that transformation is given by
$I^{ij}={\rm STF}_{ij}[I_{\rm EW}^{ij}+{1\over 21}
(11I_{\rm EW}^{ijkk}-12I_{\rm EW}^{kijk}+4I_{\rm EW}^{kkij})
+{1\over 63}(23I_{\rm EW}^{ijkkll}
-32I_{\rm EW}^{kijkll}+10I_{\rm EW}^{kkijll} +2I_{\rm EW}^{klklij})]$.
We find that those STF moments agree exactly with the ``source''
moments of BDI, e.g. Eq.~(\ref{quadrupole}).  Note that the
formal EW approach misses the tail effects in the waveform
(see~(\ref{tail})).  They must be added separately.

\acknowledgments

This work is supported in part by CNRS, the NSF
under Grant No. 92-22902 (Washington University),
and NASA under Grants No. NAGW 3874 (Washington University),
NAGW 2936 (Northwestern University) and NAGW 2897 (Caltech).
BRI acknowledges the hospitality of IHES.

\begin{table}
\caption{Contributions to the accumulated number $\cal N$ of
gravitational-wave cycles in a LIGO/VIRGO-type detector.
Frequency entering the bandwidth is 10~Hz (seismic limit);
frequency leaving the detector is 1000~Hz for 2 neutron
stars (photon shot noise), and $\sim$~360~Hz and $\sim$~190~Hz
for the two cases involving black-holes (innermost stable orbit).
Spin parameters $\beta$ and $\sigma$ are defined in the text.
Numbers in parentheses indicate contribution of finite-mass
($\eta$) effects.}
\begin{tabular}{lccc}
&$2\times 1.4M_\odot$&$10M_\odot+1.4M_\odot$&$2\times 10M_\odot$\\
\hline
Newtonian&16,050&3580&600\\
First PN&439(104)&212(26)&59(14)\\
Tail&$-$208&$-$180&$-$51\\
Spin-orbit&17$\beta$ &14$\beta$ &4$\beta$\\
Second PN&9(3)&10(2)&4(1)\\
Spin-spin&$-2\sigma$&$-3\sigma$&$-\sigma$\\
\end{tabular}
\label{table1}
\end{table}


\begin{references}
\bibitem{spincomment}
In the presence of spin-dependent interactions, a circular orbit
only exists as an approximation; the formulas presented here are
averaged over one orbit.  The only other effects that might contribute
are finite-size, quadrupole effects;  these would be significant
only for fast-rotating compact bodies, for which they scale as
$(R_0/r)^2 \sim (m/r)^2$, {\it i.e.} 2PN relative to Newtonian order,
where $R_0 \sim m$ is the size of the compact body.  However they
enter via the equations of motion, not via the radiation field at
2PN order.
\end{references}
\end{document}